\documentclass[%
 reprint,
 amsmath,amssymb,
 aps,
]{revtex4-2}

\usepackage{xcolor}

\usepackage{graphicx}
\usepackage{dcolumn}
\usepackage{bm}

\begin{document}

\preprint{APS/123-QED}

\title{Rectification from band-gap oscillation}

\author{Anwesha Chattopadhyay}
 \affiliation{Department of Physics, School of Mathematical Sciences,
Ramakrishna Mission Vivekananda Educational and Research Institute, Belur, Howrah 711202, India.}
\date{\today}
\begin{abstract}
We study a one-dimensional chain of {identical atoms with two electronic orbitals and two electrons per atom}, subject to an external oscillating pressure that periodically modulates the lattice spacing. This leads to time-dependent intra- and inter-orbital hopping amplitudes. In the tight-binding limit with weak inter-orbital hopping, the system exhibits two electronic bands separated by an oscillating indirect band-gap. By tuning hopping amplitudes and orbital energies, a periodic metal-insulator transition emerges in the half-filled system. When the frequency of this transition matches that of an external alternating electric field, the system undergoes half-wave rectification: it behaves as a conductor during one half-cycle and as an insulator during the other. This dynamic switching enables directional current flow and remains robust under small to intermediate onsite electron-electron interactions. 
\end{abstract}

\maketitle


\section{Introduction}

Metal-insulator transitions (MIT)~\cite{MetalInsulator,Nano1,Nano2,Nano3,Elec} in different systems have been studied extensively in the community for the last few decades. A metal-insulator transition can be induced by thermal fluctuations, i.e., by varying temperature~\cite{Temperature}, by doping~\cite{Doping}, by varying pressure~\cite{review1,review2, pressure_recent,pressure_bilayer,pressure_induced,pressure_abinitio,strain}, by tuning magnetic field~\cite{magnetic_1,magnetic_2}, by  tuning disorder~\cite{anderson}, by structural deformations~\cite{structural} and by quantum fluctuations at temperature $T=0$~\cite{Quantum}. Out of all these possible routes, applying pressure to induce a metal-insulator transition has a simple physics behind it. Within the paradigm of band theory where due to the quasiparticle picture of effective degrees of freedom one can ignore the electron-electron correlation effects, compressing the system decreases the lattice spacing and hence increases the overlap, or equivalently the hopping between the atomic orbitals and the bandwidth of the bands. This can make a band insulator at $T=0$, a band metal in suitable systems. However, there are a wide range of systems where electron-electron correlation is strong and cannot be neglected. For example, $V_{2}O_{3}$~\cite{V2O3_1,V2O3_2} in the undoped state is a Mott insulator~\cite{Mott_1,Mott_2} which can undergo pressure induced transition to a metallic state~\cite{V2O3_3}. This happens because compressive pressure can effectively reduce the electron-electron correlation to hopping amplitude ratio and make the system metallic. Pressure has also been found to induce superconductivity in some systems~\cite{Superconductivity, WTe2_1,WTe2_2,WTe2_3}.

In this paper, we are interested specifically in the pressure dependence of metal-insulator transition. We consider a half-filled monoatomic chain consisting of two orbitals per atom having length $L$ which is being periodically stretched and compressed using an external oscillating pressurizer. This implies that the lattice separation $a$ is also a periodically oscillating function of time and hence the hopping amplitude $t$ also carries out an out of phase oscillation as compared to the lattice separation. If $t$ oscillates about the metal-insulator transition point, then it enters the metallic and band insulator phases periodically. This means this system behaves as a two-level resistor with resistivity values $0$ and $\infty$ in periodic cycles of the metal-insulator transition frequency $\omega$. This has significant technological implications, which will be detailed in the next paragraph.

The periodic metal-insulator transitions can potentially be used for voltage rectification purposes. We know that center-tapped transformer~\cite{Transformer}    circuits use the rectification properties of $p-n$ junction diodes~\cite{Transformer} for converting alternating currents to direct currents which is required for running our electrical appliances. A $p-n$ junction is made from a $p$-type semiconductor which is basically hole doped and an $n$-type semiconductor which is electron doped. At their junction, a depletion region is formed which decreases in width when the device is forward biased as opposed to the increasing width when the device is reverse biased. Ideally, therefore the device has very low resistance when forward biased and very high resistance when reverse biased and hence a single diode acts as a half-wave rectifier to an alternating voltage. We propose that a system periodically stretched and compressed with frequency $\omega$ can serve as a half-wave rectifier for an alternating voltage with same frequency $\omega$. In the metallic phase cycle, the system conducts while in the band insulator cycle the system does not conduct any electricity. Therefore, $p-n$ junction diodes in rectification circuits can be replaced by such systems which can harbor a periodic metal-insulator transition in response to periodic pressure. {The main motivation hence is to search for rectification mechanisms beyond conventional p–n junction diodes where nanoscale systems with strain-induced, dynamic metal–insulator transitions can enable rectification of AC signals. While pressure-driven metal–insulator transitions are well established, utilizing periodic, dynamic transitions for rectification is, to our knowledge, an entirely new concept. Here, rectification is governed by the global conductivity of the material rather than the local properties of a junction.
  There are many potential candidate materials which can be used in building these nano-rectifiers working on this principle. Notable among these are transition metal dichalcogenides like $TiS_{2}, MoS_{2}$, which show a semiconductor to semimetal transition under pressure~\cite{TiS2,MoS2_1,MoS2_2} and $WTe_{2}$ which undergoes a semi-metal to superconductor transition~\cite{WTe2_1,WTe2_2,WTe2_3}. Quasi-one-dimensional Peierls distorted chains~\cite{peierls_1,peierls_2,peierls_3} like trichalcogenides (e.g., $TaSe_3$, $NbSe_3$) or organic charge-transfer salts ~\cite{org_cts} (e.g., TTF-TCNQ) have a direct band-gap which is proportional to the difference in alternate hopping amplitudes. Periodic modulation of the bond lengths can in principle lead to a band-gap closing and opening dynamically. Mott gaps in correlated oxides are also susceptible to strain and lead to insulator to metal transition with increasing pressure~\cite{mott_MIT_1,mott_MIT_2} since the effective Hubbard interaction, $U/t$ reduces due to increase in hopping amplitude, $t$ as a consequence of increase in overlap with pressure. }

{Correlation effects can significantly alter or modify the physics of the problem.} We, therefore, also include electron-electron correlation effects to test whether our proposal of rectification is robust enough. For values of interaction strengths comparable or little larger than the hopping amplitudes, the periodic transition between an insulator and a metal remains robust if we also consider a finite oscillating inter-orbital hopping amplitude comparable in strength to the oscillating intra-orbital hopping amplitudes. For low and  intermediate interaction strengths, a dynamic phase transition between a band insulator/Mott insulator and a ferromagnetic metal is observed. The  metallic regime also hosts interesting variation in orbital magnetization with time which has potential applications in the field of spintronics. Further increasing the interaction strength makes the system a Mott insulator whose positive band-gap nevertheless oscillates in time. Hence, rectification from a monoatomic two-orbital oscillating chain at half-filling is still robust in materials where electron-electron interaction is weak to intermediate in strength. While this is true at temperature $T=0$, at finite temperatures positive band-gap oscillations for higher values of interaction strength can be used for rectification purposes as well.

Even though we take a simple model of a half-filled monoatomic chain of oscillating length with two orbitals per atom and study it both in the absence and presence of electron-electron interactions, our conclusions are quite generic and hold for a wide range of models hosting rich phase diagrams. For example, the half-filled Ionic Hubbard model (IHM)~\cite{massimbalanceIHM,BagIHM,SCIHM} in the presence of nearest and next nearest neighbour hopping amplitudes has interesting metallic and insulating phases in the interaction-onsite potential parameter space. Oscillating hopping amplitudes means that effective strength of these parameters also oscillate with time. If the oscillation can be made about the metal-insulator transition lines, then half-wave rectification is in principle also possible for this model and in general for any model which shows metal-insulator transitions in the effective parameter space.

   In section~\ref{Model}, we introduce the model and it's basic features including the band dispersion calculations in the absence of electron-electron interactions. In section~\ref{Rectification} we introduce the technological aspect of this phenomena which discusses half-wave rectification due to periodic metal-insulator transition. In section~\ref{U}, we discuss the effects of electron-electron interaction in details. Finally, we conclude in section~\ref{Conclusion}.

\begin{figure*}[ht!]
     \begin{center}
\includegraphics[height=5.0cm,trim={0 5.8cm 0cm 1cm},clip]{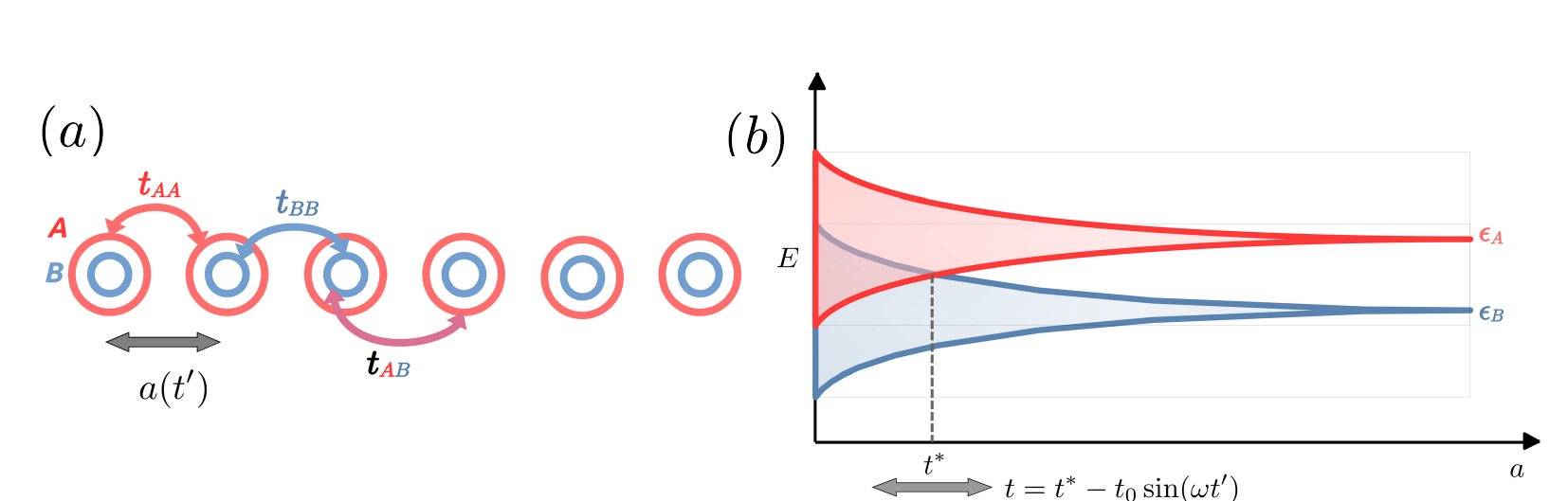}
\end{center}
   \caption{(a) Schematic shows a tight binding chain of atoms with two orbitals $A$ and $B$ on each atom. Electron hops from one atom to its nearest neighbor within same orbitals (intra) and between different orbitals (inter). The lattice spacing oscillates with time $t'$ as the length of the chain is periodically stretched and compressed. (b) Schematic shows the increasing bandwidth of the two bands with decreasing lattice spacing or increasing hopping amplitude in the static case. For simplicity the intra-orbital hopping amplitudes are considered equal in this figure with the assumption that inter-orbital hopping is vanishingly small. Oscillation of $t$ about $t^{*}$, the metal-insulator transition point, gives periodic oscillation of the band-gap, which means the system becomes metallic and insulating periodically at half-filling.  }{\label{fig1}}
\end{figure*}

\section{Model}\label{Model}

We consider a one-dimensional atomic chain made up of $N$ atoms each having two atomic orbitals $A$, $B$ with energies $\epsilon_{A}$ and $\epsilon_{B}$. The chain is being periodically stretched and compressed with a frequency $\omega$ such that the length of the chain as a function of time $t'$ is $L(t')=L^{*}+L_{0}\sin(\omega t')$, where $L^{*}> L_{0}$. This means the inter-atomic lattice spacing $a(t')$ also has the same periodic dependence $a(t')=a^{*}+a_{0}\sin(\omega t)$, where $a^{*} > a_{0}$. We will consider periodic boundary condition here, so that the system is actually a ring with equispaced atoms which is  being radially stretched and compressed. Hence the first Brillouin zone ($FBZ$) $[-\pi/a(t'),\pi/a(t'))$ will also oscillate periodically with time with the spacing between discrete crystal momentum values being $\Delta k=2\pi/L(t')$. Therefore, when the length is stretched the $k$ values come closer while the $FBZ$ also shrinks. On the other hand when the length is compressed the $k$ values move farther while the $FBZ$ also expands.

We consider both intra and inter orbital hopping between neighboring atoms, the hopping amplitudes of which we call $t_{AA},t_{BB}$ and $t_{AB}$ respectively. As the hopping amplitude is inversely proportional to the inter atomic spacing, its time dependence goes like $t_{\alpha\beta}(t')=t_{\alpha\beta}^{*}-{t_{\alpha\beta0}}\sin(\omega t')$ ($\alpha,\beta \in A,B$) as compared to $a(t')$ which goes like  $a(t')=a^{*}+a_{0}\sin(\omega t')$.  Within the tight binding approximation, the electron is tightly bound to the nuclei and we introduce a weak hopping between the neighboring atoms, which is time dependent here. Therefore, we take the following linear combination of orthonormal atomic orbitals as our ansatz solution of the wave function of the Hamiltonian,

\begin{equation}
    |\psi\rangle=\sum_{n=0}^{N-1} \phi_{n}^{A}|nA\rangle+\phi_{n}^{B}|nB\rangle.
\end{equation}

Putting this in the Schr$\ddot{o}$dinger equation and taking inner product with $\langle mA|$ and $\langle mB|$ we get,

\begin{align}
&\sum_{\alpha\in A,B}-t_{A\alpha}(\phi_{m-1\alpha}+\phi_{m+1\alpha})=(E-\epsilon_{A})\phi_{mA}\label{eqn2}\\
&\sum_{\alpha\in A,B}-t_{B\alpha}(\phi_{m-1\alpha}+\phi_{m+1\alpha})=(E-\epsilon_{B})\phi_{mB}\label{eqn3}
\end{align}

where $t_{AB}=t_{BA}$. Substituting $\phi_{mA}=\mathcal {A}e^{ikma}$ and $\phi_{mB}=\mathcal{B} e^{ikma}$ in Eq~\ref{eqn2},\ref{eqn3} we get,

\begin{equation}
    \begin{pmatrix}
       \epsilon_{A}-2t_{AA}\cos(ka) & -2t_{AB}\cos(ka)\\
        -2t_{AB}\cos(ka) & \epsilon_{B}-2t_{BB}\cos(ka)
    \end{pmatrix}\begin{pmatrix}
       \mathcal{A}\\
        \mathcal{B}   \end{pmatrix}=E\begin{pmatrix}
        \mathcal{A}\\
        \mathcal{B}.
    \end{pmatrix}
\end{equation}
Solving the characteristic equation gives us the dispersion relations for the two bands which are,

\begin{equation}
    E_{\pm}=\dfrac{(X+Y)\pm \sqrt{(X-Y)^{2}+16t_{AB}^{2}\cos^{2}(ka)}}{2},
\end{equation}

where $X=\epsilon_{A}-2t_{AA}\cos(ka)$ and $Y=\epsilon_{B}-2t_{BB}\cos(ka)$. While inter-orbital hopping tries to open a direct band-gap between the two bands , i.e., the lower band maxima and upper band minima occur at the same $k$ point, the intra-orbital hoppings favour an indirect band-gap where there is a difference in the crystal momentum of the lower band maxima and upper band minima. It is an interesting observation that tuning the strengths of $t_{AA},t_{BB}$ and $t_{AB}$ out of phase with time can give rise to oscillatory metal-insulator transitions at half-filling for suitable choice of their values at $t'=0$. However, we would not elaborate on this since it seems physically unrealistic to have out of phase oscillations of hopping parameters in a system which is being periodically stretched and compressed. Physically, all the hopping parameters should oscillate in phase. Here, we would fix the value of $t_{AB}$ at $t'=0$ to a low value or to zero and study the periodic metal-insulator transition due to periodically oscillating intra-orbital hopping $t_{AA}$, $t_{BB}$. How this happens will be elaborated in the following paragraph. It is to be noted that the sole purpose of retaining the $t_{AB}$ term above is because in an experimental sample we will in principle have this term present.

Fig~\ref{fig1}$(a)$ shows the tight binding chain with two orbitals per atom with orbital energies $\epsilon_{A}$ and $\epsilon_{B}$ such that $\epsilon_{A}>\epsilon_{B}$. We also consider intra-orbital hoppings $t_{AA},t_{BB}\gg t_{AB}$. Let us first understand the static case. In the limit of vanishing $t_{AB}$, we basically get two decoupled bands whose bandwidth is a decreasing function of lattice separation and an increasing function of hopping amplitude, as shown in Fig.~\ref{fig1}$(b)$. There is a point $t^{*}$ corresponding to $a^{*}$ which marks the onset of metallic behavior for a half-filled system from insulating behavior (at temperatures where thermal fluctuations are not able to provide the minimum single particle excitation energy)  as we decrease lattice separation. This is an example of metal-insulator transition. Now if we stretch and compress our system about $a^{*}$ periodically such that there is an out of phase oscillation in $t$ about $t^{*}$, we pass through the metal-insulator transition periodically , i.e., in one half of the cycle the system is an insulator whereas in the other half of the cycle the system is a metal.  In the band insulator phase, the lower band is completely filled which means there are $2N$ electrons in the system and at $T=0$ or low temperatures, a small electric field cannot provide the minimum energy excitation $\Delta(t')$ and hence cannot conduct electricity. When the same system undergoes a transition to the metallic phase, the bands overlap and we have partially filled bands which would respond to a small electric field by conducting current at half-filling. 

It is worthwhile to mention here that the two orbital tight binding chain described above is mathematically equivalent to the two-leg fluxless Creutz ladder~\cite{Creutz1,Creutz2} without any hopping along the rungs of the ladder. Here the two orbitals correspond to the two legs of the ladder and the hopping amplitudes, $t_{AA,BB}$ correspond to the hopping amplitudes along the legs and hopping amplitude, $t_{AB}$ corresponds to the diagonal inter-leg hopping. The inter-orbital same site hopping which corresponds to the hopping along the rungs of the ladder was not considered in our analysis. Therefore, the above analysis of introducing time-dependent hopping amplitudes stands equally well for this model. In this context, it will be interesting to study this model, specially to understand its topological implications. This is reserved for future work.

In the following section, we will see how this oscillation between metallic and insulating phases is used for rectification of alternating electric fields. 

\section{Rectification}\label{Rectification}

We all know that a semiconductor device consisting of a junction of a $p-$type and an $n-$type doped semiconductor, also known as a diode has very small resistance when it is forward biased and very high resistance when it is reverse biased. A $p-$type semiconductor like silicon doped with boron is the one where the majority charge carriers are holes whereas $n-$type semiconductor like silicon doped with phosphorous is the one where the majority charge carriers are electrons. The depletion region barrier at the junction of two types of semiconductors is reduced when the system is forward biased enabling a flow of current whereas when reverse biased the barrier height increases and hinders current flow. So it acts like a switch enabling current flow in one direction and inhibiting current flow if polarity is reversed. Thus, if we apply an alternating electric field across its terminal, the $p-n$ junction diode will enable current flow in phase with the voltage during one cycle whereas in the other half of the cycle it will give no current response. This is known as half-wave rectification.

\begin{figure}[ht!]
     \begin{center}
\includegraphics[height=5.3cm,trim={2.0cm 0 0cm 0cm},clip]{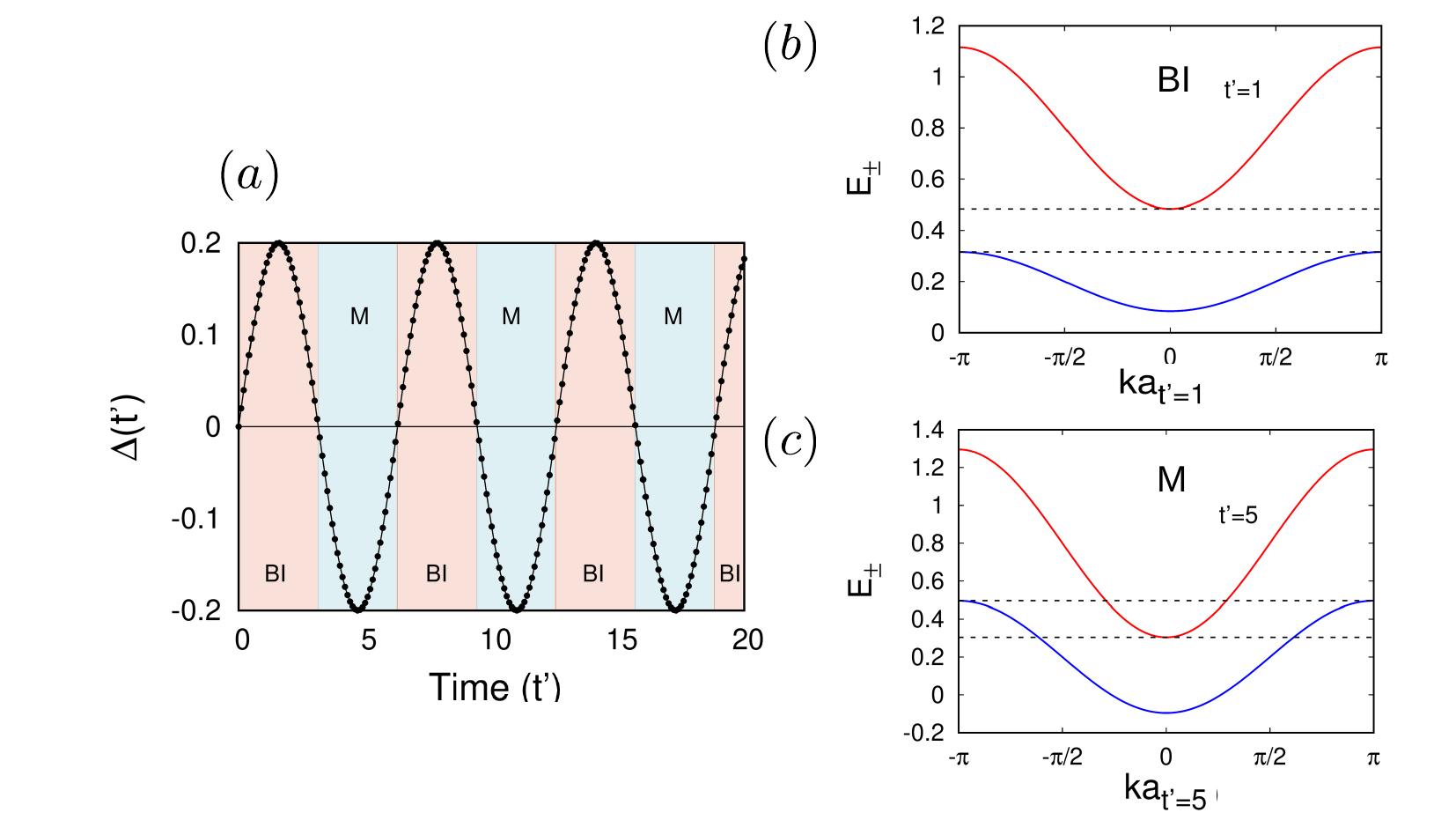}
\end{center}
   \caption{(a) Figure shows plot of the band-gap $\Delta(t')$ as a function of time $t'$ for $\epsilon_{A}=0.8,\epsilon_{B}=0.2$ and hopping amplitude oscillation of $0.05$ around $t_{AA}^{*}=0.2,t_{BB}^{*}=0.1$. For $\Delta(t')>0$, the system is a band insulator (shaded as red) whereas for $\Delta(t')<0$, the system is a metal (shaded as blue) at half-filling. (b),(c) show the band dispersions at $t'=1$ and $t'=5$ where the system is a band insulator and a metal respectively at half-filling.}{\label{fig2}}
\end{figure}

We propose a novel mechanism of half-wave rectification through periodic metal-insulator transition in a chain of atoms, the length of  which is being stretched and compressed periodically. In the previous section we have shown that if the intra-orbital hopping parameters oscillate about $t^{*}$ as shown in Fig~\ref{fig1}$(b)$ and the system is half-filled, then the system periodically enters a metallic phase and a band insulator phase. To characterize these two phases we define the band-gap $\Delta(t')$ which is equal to $\epsilon_{A}-\epsilon_{B}-2(t_{BB}+t_{AA})$ for $t_{AB}=0$ in our model. The term band $gap$ is a little misleading when $\Delta(t')<0$ as it is more appropriate to call it as an $overlap$ instead of gap. Nevertheless, to keep uniformity in the notation we will refer to band overlap as negative band-gap. For the band insulator phase, $\Delta(t')>0$ , i.e., the valence band maxima lies below the conduction band minima whereas for the metallic phase, $\Delta(t')<0$ when the bands overlap and the conduction band minima lies below the valence band maxima. Fig~\ref{fig2}$(a)$ shows the plot of $\Delta(t')$ as function of the time $t'$ for the hopping parameters $t_{AA}^{*}=0.2,t_{BB}^{*}=0.1, t_{{AA0}}= t_{{BB0}}=0.05$ ($t_{AB}$ has been taken to be zero) and onsite orbital energies $\epsilon_{A}=0.8,\epsilon_{B}=0.2$. These parameters correspond to $\Delta(t'=0)=0$ when the valence band maximum at $ka(t'=0)=\pi$ just equals to the conduction band minimum at $ka(t'=0)=0$. The time period of this oscillation is $T=2\pi/\omega$ where $\omega=1$. At half-filling, in the red shaded regions the system is a band insulator (BI) whereas in the blue shaded regions the system is metal (M). Fig~\ref{fig2}$(b),(c)$ shows the band dispersions at $t'=1$ and $t'=5$ where the system is a band insulator and a metal respectively at half-filling. 

\begin{figure}[ht!]
     \begin{center}
\includegraphics[height=3.8cm,trim={0 3.5cm 0cm 0cm},clip]{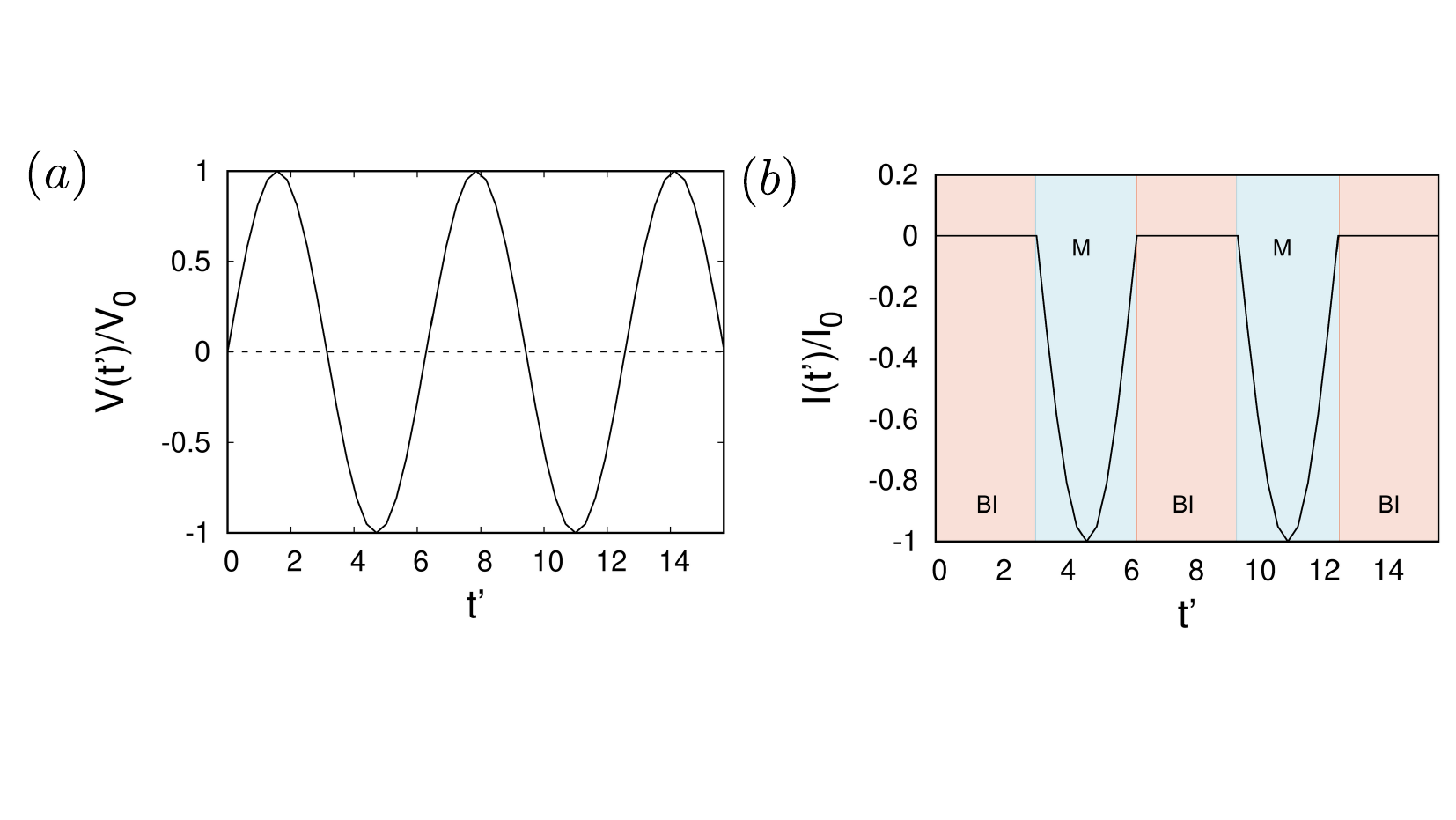}
\end{center}
   \caption{(a)-(b) An external alternating voltage $V(t')=V_{0}\sin(\omega t')$ with $\omega=1$ , i.e., having the same frequency as that of the oscillating pressure subjected on the chain is half-wave rectified due to a periodic metal-insulator transition at half-filling. Here $I(t')$ is the current in the system due to the external electric field.}{\label{fig3}}
\end{figure}

 Now if we consider an alternating electric field or equivalently a voltage which has the same time period as the periodic pressure on the chain then the half-filled chain would conduct in the half-cycle where it is a metal and will not conduct in the other half-cycle. This is shown in Fig~\ref{fig3}$(a),(b)$. So, just like a $p-n$ junction diode it acts like a half-wave rectifier, only the principles of operation are very different. However, at finite but low temperatures of operation, an asymmetry in the metallic and insulator regions is expected because for very small positive band-gaps, thermal excitations can make the system metallic.

An important point to ponder about is whether quantum adiabaticity holds in each cycle of the metal-insulator oscillation or not. The band structure evolves dynamically according to the Hamiltonian but the electronic degrees of freedom require some time to follow the dynamics. The band structure variation should be slow enough so that the electrons remain in the instantaneous eigenstates of the Hamiltonian. If the band-gap varies like $\Delta(t')=\Delta_{0}\sin(\omega t')$, then  the band-gap closes at $t'=n\pi/\omega$ where $n\in Z^{+}$ and the Landau Zener adiabaticity condition $\omega \ll \Delta(t')$ fails near the vicinity of metal-insulator transitions. This is a truly non-adiabatic regime where there will be non-trivial effects. In the paper, $\omega$ has been fixed to be unity for convenience but it can be made smaller without changing the qualitative features of the results. This will strengthen the adiabaticity argument in most of the time domain of the cycle apart from a narrow regime around $\Delta(n\pi/\omega)=0$ where the $I-V$ characteristics will presumably be noisy. Current may leak into the nominally insulating regions creating asymmetry in the metallic and insulating regions as well as there may be hysteresis effects within this narrow window. However, this smearing in the $I-V$ characteristics in this narrow regime does not invalidate the periodic oscillation between metallic and insulating phases in every cycle, making it suitable for rectification processes. Moreover, at finite temperatures of operation, there is no need of actual gap closing for rectification; there can be a $\Delta_{min}>0$ in the spectrum and the positive band-gap can oscillate with the oscillating lattice separation, such that there is a critical band-gap $\Delta_{c}(T)$ below which we can get a thermally assisted metal and above which we have an insulator.

\section{Effect of electron-electron interaction, $U$}\label{U}

In the previous discussion we did not take into account the effect of electron-electron interaction which as we shall see plays a crucial role. In this section, we add onsite repulsive Hubbard terms on both orbitals in the tight binding model described previously and treat $U\geq t_{\alpha\beta}, \alpha,\beta \in A,B$  at the Hartree-Fock mean field level. For completeness, we will also consider a finite oscillating inter-orbital hopping amplitude, $t_{AB}(t')$ in this discussion. The model, in the second quantized notation, we will be considering is as follows,

\begin{align}\label{orbitalHubbard}
    \mathcal{H}=&\sum_{\alpha\beta\in A,B}\sum_{<ij>,\sigma}(-t_{\alpha\beta}(t')c_{i\alpha\sigma}^{\dagger}c_{j\beta\sigma} +h.c.)+\sum_{i,\alpha}\epsilon_{\alpha}n_{i\alpha}\nonumber\\
   & +U\sum_{\alpha,i}n_{i\alpha\uparrow}n_{i\alpha\downarrow}
\end{align}

where $t_{AB}=t_{BA}$ and should be counted once in the summation. We do an unrestricted Hartree-Fock mean field theory to get an effective quadratic Hamiltonian which we diagonalize using a standard canonical transformation, the details of which are provided in Appendix $A$. The effective dispersions are as follows,

\begin{align}\label{bands}
   \lambda^{1,2}_{\sigma}=&\bigg[\dfrac{U(1-\sigma m_{f})}{2}+\dfrac{(\epsilon_{A}+\epsilon_{B})}{2}-\mu-\dfrac{1}{2}\sum_{\alpha\in A,B}t_{\alpha\alpha}(t')\gamma_{k} \bigg]\nonumber\\
   &\mp \sqrt{\tilde{\Delta}_{\sigma}^{2}+t_{AB}^{2}(t')\gamma_{k}^{2}} 
\end{align}

{where $\lambda^{1,2}_{\sigma}$ are the effective valence and conduction bands respectively which comes in two spin-polarized variants, $\sigma=\uparrow,\downarrow$. On the right,} $\tilde{\Delta}_{\sigma}=(U(\delta-\sigma m_{s})+(\epsilon_{A}-\epsilon_{B})-(t_{AA}-t_{BB})\gamma_{k})/2$   and $m_{s,f}$ are the staggered and uniform magnetizations defined as $(m_{A}\mp m_{B})/2$, where $m_{A,B}$ are the orbital magnetizations of $A$ and $B$ orbitals. Also, $\delta$ is the density difference between the orbitals defined as $(n_{A}-n_{B})/2$ where $n_{A,B}$ are the electron densities of the $A$ and $B$ orbitals. We solve the Hamiltonian for overall density, $(n_{A}+n_{B})/2=1$. Here, $\gamma_{k}=2\cos(ka)$.  

\begin{figure}[ht!]
     \begin{center}
\includegraphics[height=5cm,trim={0 0cm 0cm 0cm},clip]{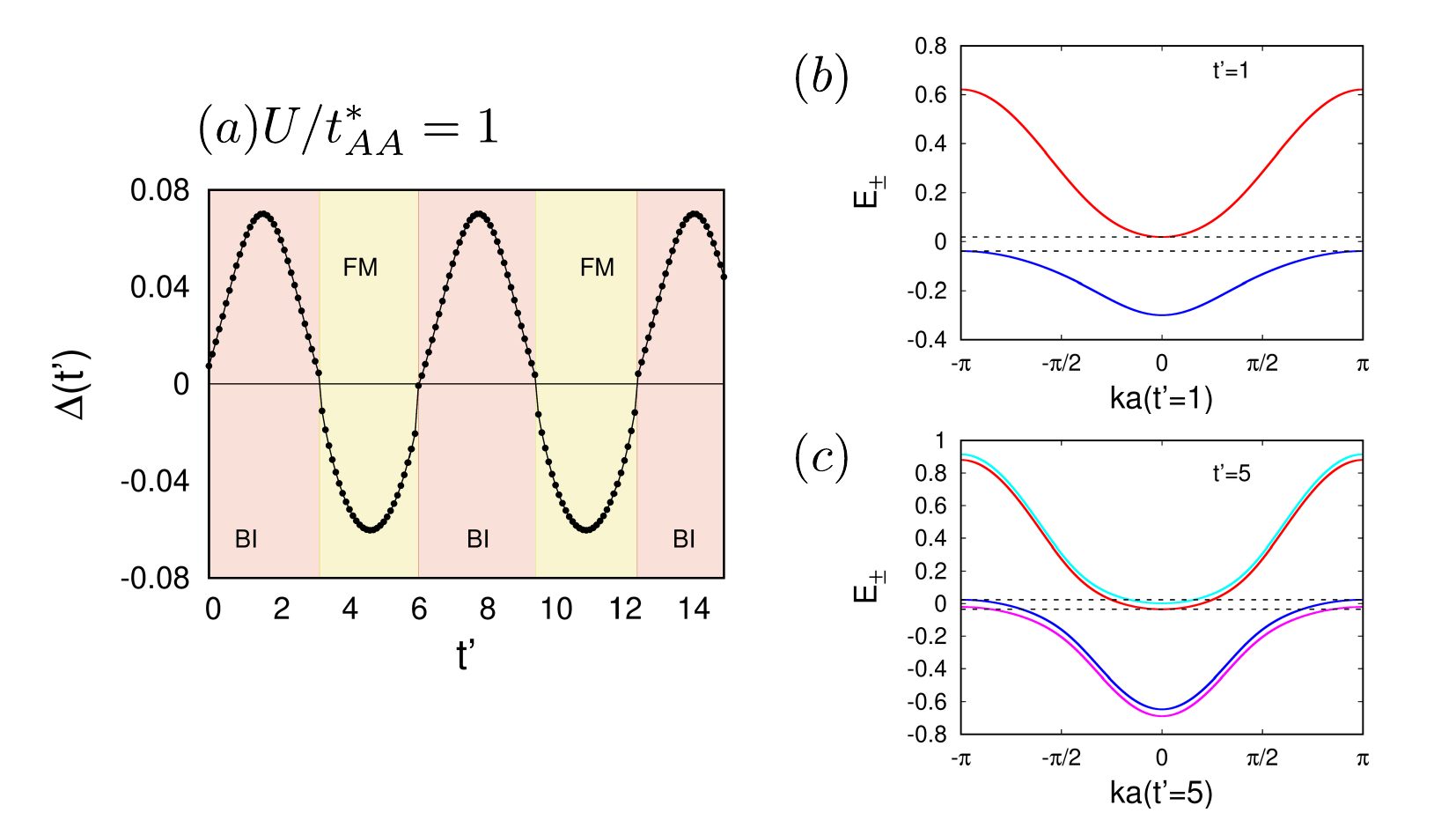}\\ \includegraphics[height=5cm,trim={0 0cm 0cm 0cm},clip]{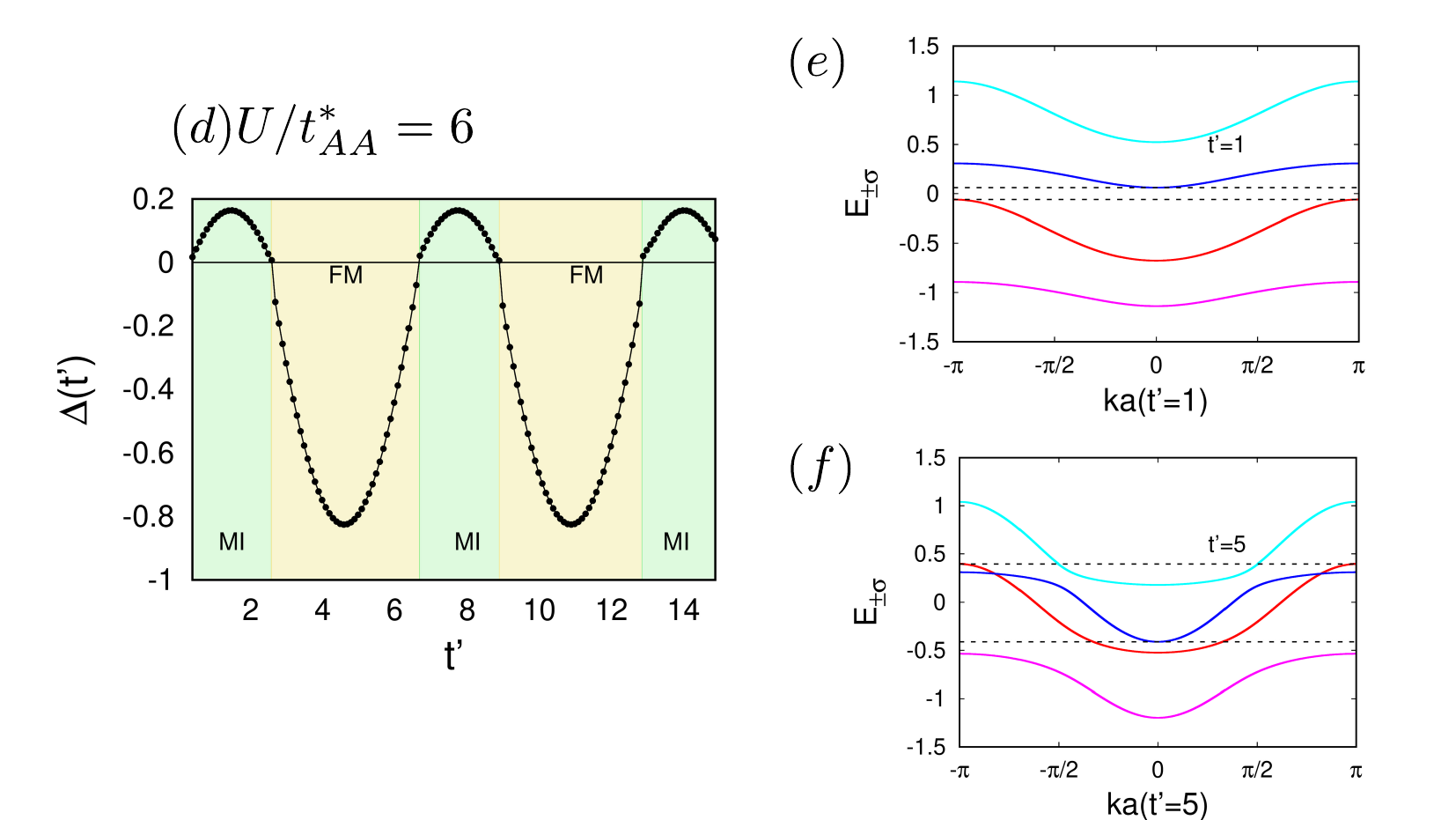}\\  \includegraphics[height=5cm,trim={0 0cm 0cm 0cm},clip]{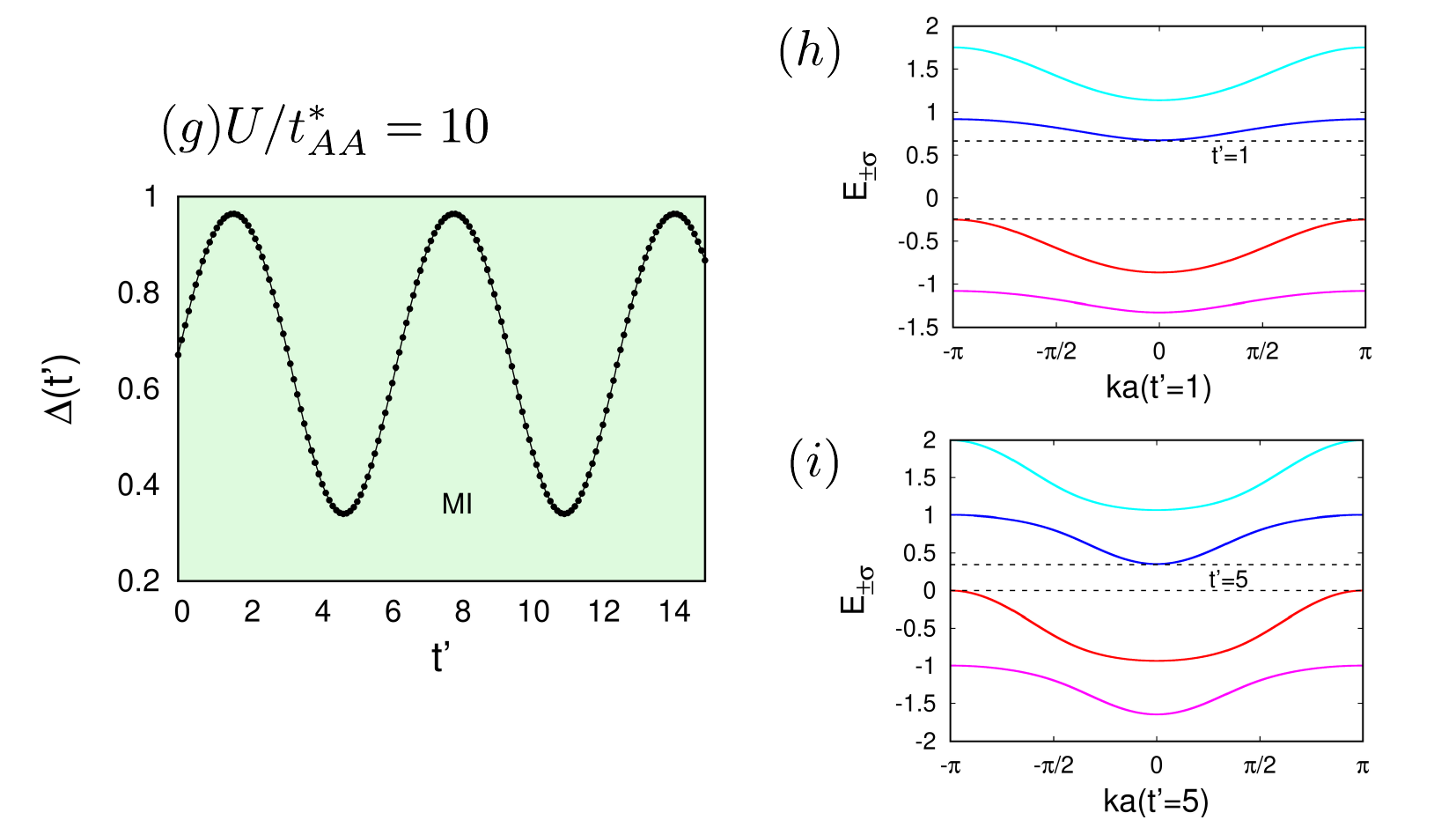}
\end{center}
   \caption{Panels $(a)-(c)$ show the oscillation of band-gap $\Delta(t')$ for $U/t_{AA}^{*}=1$ and the band dispersions for $t'=1,5$. There is an oscillation between a band insulator (BI) and an orbital imbalanced ferromagnetic metal (FM) in this case. Panels $(d)-(f)$ show the same for $U/t_{AA}^{*}=6$ where there is a dynamic phase transition between a Mott insulator (MI) and an orbital imbalanced ferromagnetic metal (FM). Panels $(g)-(i)$ also show the same for $U/t_{AA}^{*}=10$ where the system is always a MI but with oscillations in the positive value of the band-gap.}{\label{fig4}}
\end{figure}

\begin{figure}[ht!]
     \begin{center}
\includegraphics[height=3.8cm,trim={1cm 4cm 1cm 1cm},clip]{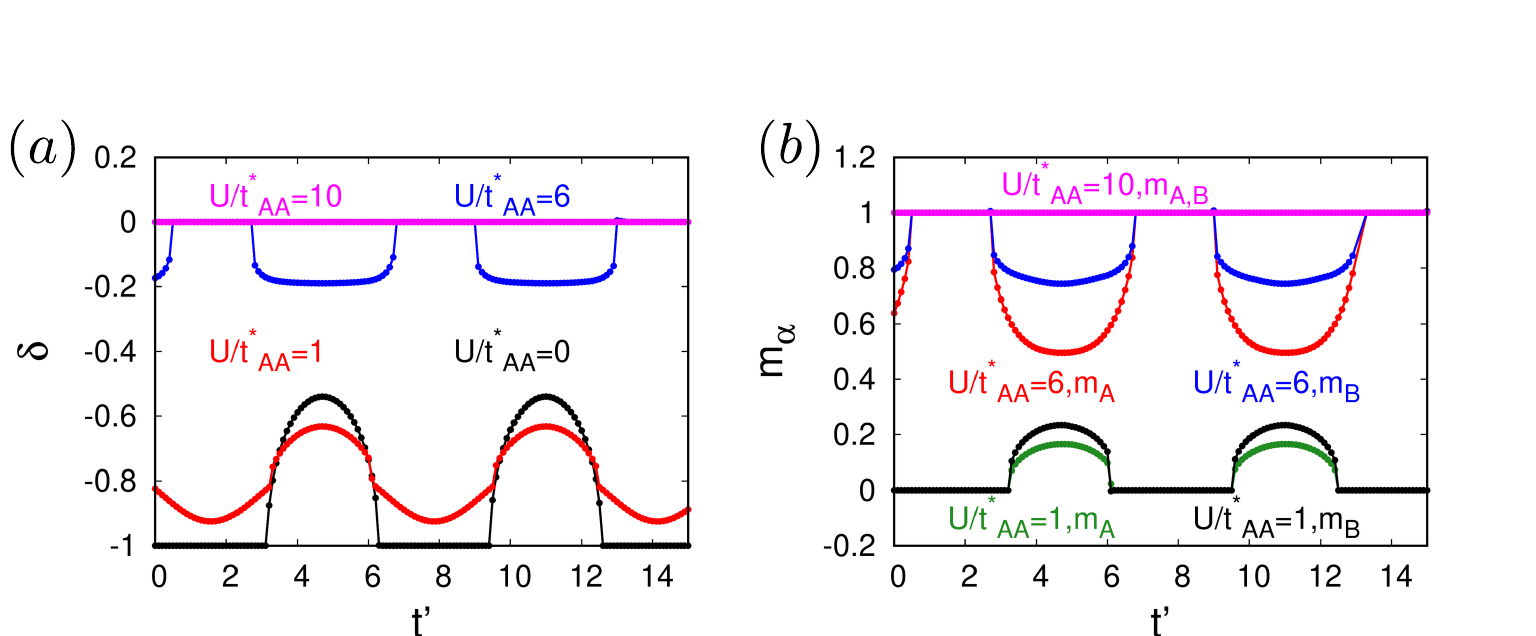}
\end{center}
   \caption{Panel $(a)$ shows the density difference, $\delta$ between the two orbitals $A$ and $B$ as a function of time, $t'$ for $U/t_{AA}^{*}=0,1,6,10$. Panel $(b)$ shows the orbital magnetization, $m_{\alpha}$ where $\alpha \in A,B$ as a function of time, $t'$ for $U/t_{AA}^{*}=1,6,10$. For $U/t_{AA}^{*}=0$, there is no orbital magnetization.}{\label{fig5}}
\end{figure}

At $U/t_{AA}^{*}=1$, the system oscillates between a band insulator (BI) with zero orbital magnetization and an orbital imbalanced ferromagnetic metallic (FM) phase where $m_{A} < m_{B}$. The polarity of the band-gap, $\Delta(t')$ oscillates with time $t'$ as can be seen in Fig~\ref{fig4}$(a)$. In panels $(b)-(c)$, we see that at $t'=1$ there is a positive band-gap in the spectrum whereas for $t'=5$, the band-gap is negative, which implies an overlap of the bands and hence a metallic phase. In the BI phase, the effective $U$ is weak and is not capable of creating spin asymmetry in the system. However, in the metallic phase $U$ creates a small spin asymmetry in the spectrum which creates magnetization in the system because up and down spin bands are not populated/de-populated simultaneously as they are energetically non-degenerate now. The variation of orbital magnetization can be seen in Fig~\ref{fig5}$(b)$. Therefore, there is a dynamical phase transition between symmetry broken and unbroken phases which is a new concept emerging out of this simple model. Here, the orbital energy values $\epsilon_{A,B}$ and the intra-orbital hopping amplitudes, $t_{\alpha\alpha}$ where $\alpha\in A,B$ are kept the same as in the non-interacting case. The inter-orbital hopping, $t_{AB}(t')=0.1-0.05\sin(t')$ which has been kept the same for all the $U \neq 0$ cases considered.  Due to the inter-orbital hopping, the bands lose the orbital character and the bands can no longer be associated with each orbital. This is also the reason why the absolute value of the density difference, $\delta$ between the orbitals is never unity even though pairs of up and down spin degenerate bands lie below and above the Fermi level in the band insulator phase. This is seen in Fig~\ref{fig5}$(a)$. In contrast to this, at $U=0$ and $t_{AB}=0$, in the band insulator phase the absolute value of the density difference is unity which is due to the fact that the lower two spin-degenerate bands belonging to $B$ orbital are occupied and the upper ones corresponding to the $A$ orbital are empty. This can also be seen in Fig~\ref{fig5}$(a)$.

{We have seen that in the non-interacting case, inter-orbital hopping tries to open a direct band-gap which competes with the indirect band-gap in the presence of only intra-orbital hopping amplitudes. To keep the picture of metal-insulator transition clean, we assumed $t_{AB}$ to be weak. The relative magnitudes of the hopping amplitudes are determined by the symmetry of the orbitals which decides the extent of overlap.  } However, it is important to mention that at $U/t_{AA}^{*}=1$, to get a finite time domain of correlated band insulating phase, a finite value of $t_{AB}$ comparable to $t_{AA},t_{BB}$ must be included in the analysis. Since $U$ prefers single occupancy on both orbitals, in the absence of $t_{AB}$, it simply tries to push the upper bands below the Fermi level so that absolute density difference between the orbitals becomes less than unity. Introduction of oscillating $t_{AB}$, produces a positive band-gap, creating a correlated band insulator phase in specific time domains. 

At $U/t_{AA}^{*}=6$, there is a  sufficiently large spin asymmetry in the spectrum. There are now oscillating phase transitions between a Mott insulator (MI) and a ferromagnetic metal (FM) with      $m_{A} > m_{B}$ with time as can be seen in Fig~\ref{fig4}$(d)-(f)$ and Fig~\ref{fig5}$(b)$. In the MI phase, same spin polarization band pairs appear below and above the Fermi level (Fig~\ref{fig4}$(e)$) with $m_{A}=m_{B}=1$ and density difference between orbitals zero as seen in Fig~\ref{fig5}$(a)-(b)$. This means there is a spin gap in the system. In the FM phase, one of the bands of opposite spin polarity than the bands below Fermi level start crossing the Fermi level whereas one of the previously occupied bands get partially unoccupied, maintaining overall density to be unity. This can be seen in Fig~\ref{fig4}$(f)$. This creates time-varying magnetization which differs from unity in both orbitals as can be seen in Fig~\ref{fig5}$(b)$.

At a higher value of $U$, say $U/t_{AA}^{*}=10$, the spin symmetry remains broken and the $t_{AB}$ term becomes less important/negligible as compared to $U$ such that the one to one correspondence between bands and orbitals are restored. The intra-orbital hopping amplitudes now become incapable of closing the gap into a metallic phase and the system is always a  MI with oscillating positive band-gap with time (Fig~\ref{fig4}$(g)-(i)$) and with unit magnetization for both orbitals as is seen in Fig~\ref{fig5}$(b)$. This means each orbital is now singly occupied due to the dominance of electron-electron repulsive interaction over hopping which tends to delocalize electrons as opposed to $U$, which in turn means $\delta=0$ as is seen in Fig~\ref{fig5}$(a)$ . Now, two effective bands of same spin and belonging to two different orbitals lie below the Fermi level as can be seen in Fig~\ref{fig4}$(h)-(i)$. The oscillating band-gap in this case can give rise to interesting optical properties in the system, for example, the system can change color while it is being periodically stretched and compressed under the condition that $\omega_{oscillation}\ll \omega_{light}$. 

\begin{figure}[ht!]
     \begin{center}
\includegraphics[height=3.6cm,trim={1cm 4cm 1cm 1cm},clip]{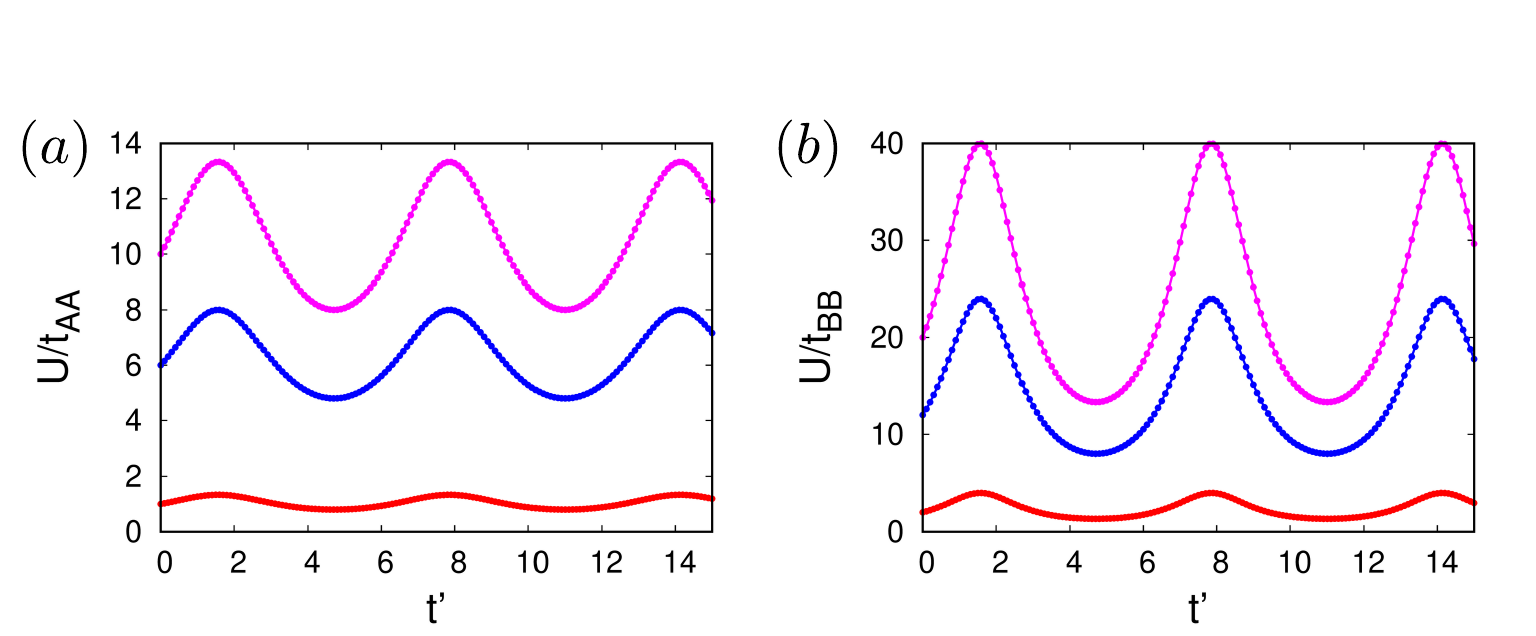}
\end{center}
   \caption{Panels $(a)-(b)$ show the variation of $U/t_{AA,BB}$ with time for $U/t_{AA}^{*}=1$ (red), $U/t_{AA}^{*}=6$ (blue) and $U/t_{AA}^{*}=10$ (magenta). $U/t_{AB}$ is the same as $U/t_{BB}$ in our calculation. }{\label{fig6}}
\end{figure}

Since the hopping amplitudes are all oscillating in phase with time in response to the periodic stretching and compression, the effective $U$ also oscillates in time as seen in Fig~\ref{fig6}. Our results are consistent with the expectation that in regions where $U/t_{\alpha\beta}$ where $\alpha,\beta \in A,B$ is small enough metallic regions appear whereas for stronger values we have insulating phases. Hence there is a dynamic phase transition in the system unless the effective $U$ remains always large, in which case it is a Mott insulator throughout.  So, we conclude that for stronger values of $U$, the system remains an insulator and cannot be used for rectification purposes until and unless hopping strength can be made quite strong, so that their ratio becomes effectively small. However for $U$ comparable or slightly greater than $t_{\alpha\beta}$ where $\alpha,\beta\in A,B$, the system still can be used as a half-wave rectifier if inter-orbital hopping is also finite, as is always present in a system. The statements we are making are true at temperature, $T=0$. However, at finite temperatures, there exist a critical band-gap , $\Delta_{c}(T)$ below which the system can have thermal excitations to the conduction band and hence is a metal. Therefore, for $\Delta(t')>\Delta_{c}(T)$, the system is an insulator whereas for $\Delta(t')<\Delta_{c}(T)$, the system is a metal. So, rectification is possible at finite temperatures even for high values of $U$. In addition at $T=0$, we observe interesting dynamical phase transitions between BI/MI and orbital imbalanced FM which can certainly have technological implications in nanoscale spintronics.  However, at finite temperatures, Mermin-Wagner theorem~\cite{MW} dictates that ordered phases due to spontaneous symmetry breaking of continuous symmetries cannot be stabilized due to dominating fluctuations in the system which destroys long range order for dimensions, $d\le 2$. Hence, the ferromagnetic phase predicted within Hartree-Fock mean field theory may not be realizable in practice for $d\le 2$.

\section{Conclusion}\label{Conclusion}

In conclusion, we have proposed a one-dimensional model of atoms in a chain whose length is being stretched and compressed periodically for example by providing oscillating pressure. As a consequence we see, that it is possible for the system to undergo repeated metal-insulator transitions which can be used for the purpose of half-wave rectification of alternating voltages.  This remains robust even in the presence of  electron-electron interaction strengths which are weak to intermediate as compared to the intra and inter orbital hopping amplitudes. Moreover, at finite temperatures, rectification is even possible for higher values of interaction strength due to the oscillation of positive band-gap. Therefore this can potentially replace $p-n$ junction diodes whose rectification properties are used in center-tapped transformer circuits to convert alternating current to direct current. Stretching reduces hopping which causes the electron-electron interaction to hopping ratio, $U/t$ to become large and the system behaves like a Mott insulator at half-filling. While if $t$ becomes large, i.e., if we compress the system, it will behave like a metal at half-filling.  This phenomenon is quite generic and not limited to the two-orbital Hubbard model. In the static case in the two-orbital Hubbard model, we know that within mean field theory, the system goes from a band insulating phase to a correlated (Mott) insulating phase mediated by a metallic phase with increasing interaction strength. The phase boundaries may be close but the oscillation of the effective interaction strength about the phase boundaries will not introduce any blurriness as long as we keep the oscillation amplitudes to be quite small, allowing for sharp transitions. However, we acknowledge that alternative numerical techniques like DMRG~\cite{DMRG}, Quantum Monte Carlo~\cite{QMC} can provide better insights and this is reserved as part of the future work. 

 For weak to intermediate interaction strengths, we also observed that the metallic phase is an exotic ferromagnetically ordered phase.      However, we do acknowledge that in one-dimensional systems long range order is precluded due to Mermin Wagner theorem~\cite{MW}. Therefore, we can perhaps better dub the ferromagnetic phase as a mean field magnetic instability. However, it is possible to go to dimensions $d\ge2$ and explore the possibility of such time-varying magnetizations. {Applying uniaxial strain in higher dimensional systems specially two dimensional materials like graphene~\cite{Graphene1,Graphene2, Graphene3} and TMDs~\cite{TMD} has been shown to modify the electronic band structure leading to MIT. Applying oscillating pressure on a flake can induce similar dynamic metal-insulator transitions which can again be used for rectification purposes. If electron-phonon interactions start playing a role, then it may cause dimerization which opens up a direct gap in competition with the indirect gap here. Moreover, it may cause heating and dissipation which can make the rectification effect less sharp. However, one interesting aspect is that long ranged magnetically ordered phases can be realized in higher dimensions and therefore there can be novel magnetic transitions in the system which can be used for spintronic applications. } 

 It is also worthwhile to mention here that the choice of a one-dimensional monoatomic chain was deliberate to show in a clear way the mechanism of pressure induced rectification. However, this simple model provides a generic framework and the work can be extended to systems which host metal-insulator transitions in effective parameter space like the ionic Hubbard model ~\cite{massimbalanceIHM,BagIHM,SCIHM} etc. Inclusion of terms like magnetic coupling and spin-orbit interaction relevant in real materials would perhaps be interesting from the point of view of topological transitions but  would obscure the core physical insight we want to convey. However, this analysis can serve as a foundational starting point for material specific analysis which would perhaps require $ab$ $intio$  studies which is beyond the scope of this work. Further, we have not numerically substantiated the $I-V$ characteristics with the help of non-equilibrium Green's functions since the purpose of this work is to provide a  proof-of-principle analytical work. 

{In addition, the technological prospects for this mechanism are notably significant. Strain-based nano-rectifiers can be designed where on-chip rectification circuits consists of a nanowire acting as  the diode element and whose conductivity is controlled by an external oscillating pressurizer. The generation of periodic pressure and their application on nanoscale systems can be realized by using piezoelectric actuators~\cite{Piezo1,Piezo2,Piezo3}
which work on the principle of inverse piezoelectric effect where if an alternating voltage is applied across the substrate, alternating mechanical vibrations are produced. If a nanowire or flake is deposited on such a piezoelectric substrate, it will experience oscillating strain along its length.  We can directly perform a transport measurement on these devices where, say, we measure the output current in the circuit consisting of the nanowire deposited on the vibrating piezoelectric substrate and an input AC voltage which is applied across the nanowire. Thus we can get a rectified DC signal as output in the circuit. Further, techniques like optical spectroscopy ~\cite{optical} can be used to track the oscillating band-gap, and scanning tunneling microscopy (STM)~\cite{STM} could be used to find the spatially resolved density of states, as it oscillates over time under an applied AC strain. Alternatively, oscillating pressure can also be generated using surface acoustic waves(SAW)~\cite{SAW1} which can produce such oscillating mechanical strain when the flake is placed in the path of a traveling wave. Photo-induced phonon excitation ~\cite{photo1,photo2} by using pulsed lasers is another way to realize  phonon-mediated oscillations in lattice spacing.  Also, it is worthwhile to mention energy harvesting, where mechanical vibrations (AC strain) can be directly converted into DC current in a single material device.}

Therefore to conclude, in systems where quasiparticle picture holds good and band theory works as well as in systems where electron-electron interaction plays a crucial role, periodic stretching and compression can be potentially used for rectification purposes, as in transformer circuits, replacing the widespread use of $p-n$ junction diodes.

{\it Acknowledgements}. The author would like to thank Krishnendu Sengupta for valuable suggestions.

{\section*{Appendix A}\label{Appendix}

Here, we discuss the details of the unrestricted Hartree-Fock theory used for treating the Hubbard term in the Hamiltonian described in Eq~\ref{orbitalHubbard}. We decompose the Hubbard $U$ term on each orbital in terms of spin resolved densities, $n_{i\alpha\sigma}$ where $\alpha \in A,B$ orbitals.

\begin{align}
Un_{i\alpha\uparrow}n_{i\alpha\downarrow}\approx U\langle n_{i\alpha\uparrow} \rangle n_{i\alpha\downarrow}+Un_{i\alpha\uparrow}\langle n_{i\alpha\downarrow}\rangle-U\langle n_{i\alpha\uparrow}\rangle \langle n_{i\alpha\downarrow} \rangle   
\end{align}

The effective mean field Hamiltonian, $H_{eff}$ is given by,

\begin{align}
    &H_{eff}=\sum_{k\sigma}\bigg(\dfrac{U(1+\delta-\sigma m_{A})}{2}+\epsilon_{A}-t_{AA}\gamma_{k}-\mu\bigg)c_{kA\sigma}^{\dagger}c_{kA\sigma}\nonumber\\&+\bigg(\dfrac{U(1-\delta-\sigma m_{B})}{2}+\epsilon_{B}-t_{BB}\gamma_{k}-\mu\bigg)c_{kB\sigma}^{\dagger}c_{kB\sigma}\nonumber\\&+t_{AB}\gamma_{k}(c_{kA\sigma}^{\dagger}c_{kB\sigma} +h.c.)
\end{align}

where, $\gamma_{k}=2\cos(ka)$. 

We diagonalize the Hamiltonian by using the following canonical transformation,
\begin{equation}
    \begin{aligned}
    &c_{kA\sigma}=\alpha_{k\sigma}d_{k1\sigma}+\beta_{k\sigma}d_{k2\sigma}\\ &c_{kB\sigma}=\alpha_{k\sigma}d_{k2\sigma}-\beta_{k\sigma}d_{k1\sigma}
\end{aligned}
\end{equation}

where, $\alpha_{k\sigma}^{2}=\frac{1}{2}[1-\frac{\tilde{\Delta}_{\sigma}}{\sqrt{\tilde{\Delta}_{\sigma}^{2}+t_{AB}^{2}\gamma_{k}^{2}}}]$ and $\beta_{k\sigma}^{2}=\frac{1}{2}[1+\frac{\tilde{\Delta}_{\sigma}}{\sqrt{\tilde{\Delta}_{\sigma}^{2}+t_{AB}^{2}\gamma_{k}^{2}}}]$. Here, $\tilde{\Delta}_{\sigma}=(U(\delta-\sigma m_{s})+(\epsilon_{A}-\epsilon_{B})-(t_{AA}-t_{BB})\gamma_{k})/2$.

The self-consistent equations of the spin resolved densities, $n_{\alpha\sigma}$ is given by,
\begin{equation}
    \begin{aligned}
    n_{A\sigma}=\dfrac{1}{N}\sum_{k\in FBZ}[\alpha_{k\sigma}^{2}\langle d_{k1\sigma}^{\dagger}d_{k1\sigma}\rangle+\beta_{k\sigma}^{2}\langle d_{k2\sigma}^{\dagger}d_{k2\sigma}\rangle]\\
    n_{B\sigma}=\dfrac{1}{N}\sum_{k\in FBZ}[\alpha_{k\sigma}^{2}\langle d_{k2\sigma}^{\dagger}d_{k2\sigma}\rangle+\beta_{k\sigma}^{2}\langle d_{k1\sigma}^{\dagger}d_{k1\sigma}\rangle]
\end{aligned}
\end{equation}

Here, $N$ is the total number of sites such that the length of the chain, $L(t')=Na(t')$.

The band fillings are determined by Fermi Dirac statistics. For zero temperature, the bands $\lambda^{1,2}_{\sigma}$ (given by Eq~\ref{bands} in the main text) which are below the Fermi level will be occupied. We use the spin resolved densities to construct linear combinations like orbital magnetization, $m_{\alpha}=n_{\alpha\uparrow}-n_{\alpha\downarrow}$ where $\alpha \in A,B$ and density difference between orbitals, $\delta=(n_{A}-n_{B})/2$, which are of physical interest.
}

\end{document}